\documentclass[preprint,12pt]{elsarticle}

\usepackage{graphicx}
\usepackage{amssymb}
\usepackage{amsmath}
\usepackage{array}
\biboptions{comma,square,sort&compress}

\journal{NIMA}

\begin{document}

\begin{frontmatter}

\title{Modeling crosstalk and afterpulsing in silicon photomultipliers}

\author[UCM]{J.~Rosado\corref{cor}}
\ead{jaime\_ros@fis.ucm.es}
\author[UCM]{V.M.~Aranda}
\author[UCM]{F.~Blanco}
\author[UCM]{F.~Arqueros}
\cortext[cor]{Corresponding author}
\address[UCM]{Departamento de F\'{i}sica At\'{o}mica, Molecular y Nuclear, Facultad de Ciencias F\'{i}sicas,
Universidad Complutense de Madrid, E-28040 Madrid, Spain}

\begin{abstract}
An experimental method to characterize the crosstalk and afterpulsing in silicon photomultipliers has been developed
and applied to two detectors fabricated by Hamamatsu. An analytical model of optical crosstalk that we presented in a
previous publication has been compared with new measurements, confirming our results. Progresses on a statistical model
to describe afterpulsing and delayed crosstalk are also shown and compared with preliminary experimental data.
\end{abstract}


\end{frontmatter}

\section{Introduction}
\label{sec:intro}

Silicon photomultipliers (SiPM) have excellent time and photon counting resolutions, among other qualities. However,
crosstalk and afterpulsing seriously limit the performance of these photon detectors. These phenomena are the
production of parasitic avalanches in either the pixel where a primary avalanche was triggered (afterpulsing) or a
neighboring one (crosstalk). Their main effects are to increase both the count rate and the total charge collected for
an input light pulse, e.g., from a scintillator.

\begin{figure}[t!]
\centering%
\includegraphics[width=.7\linewidth]{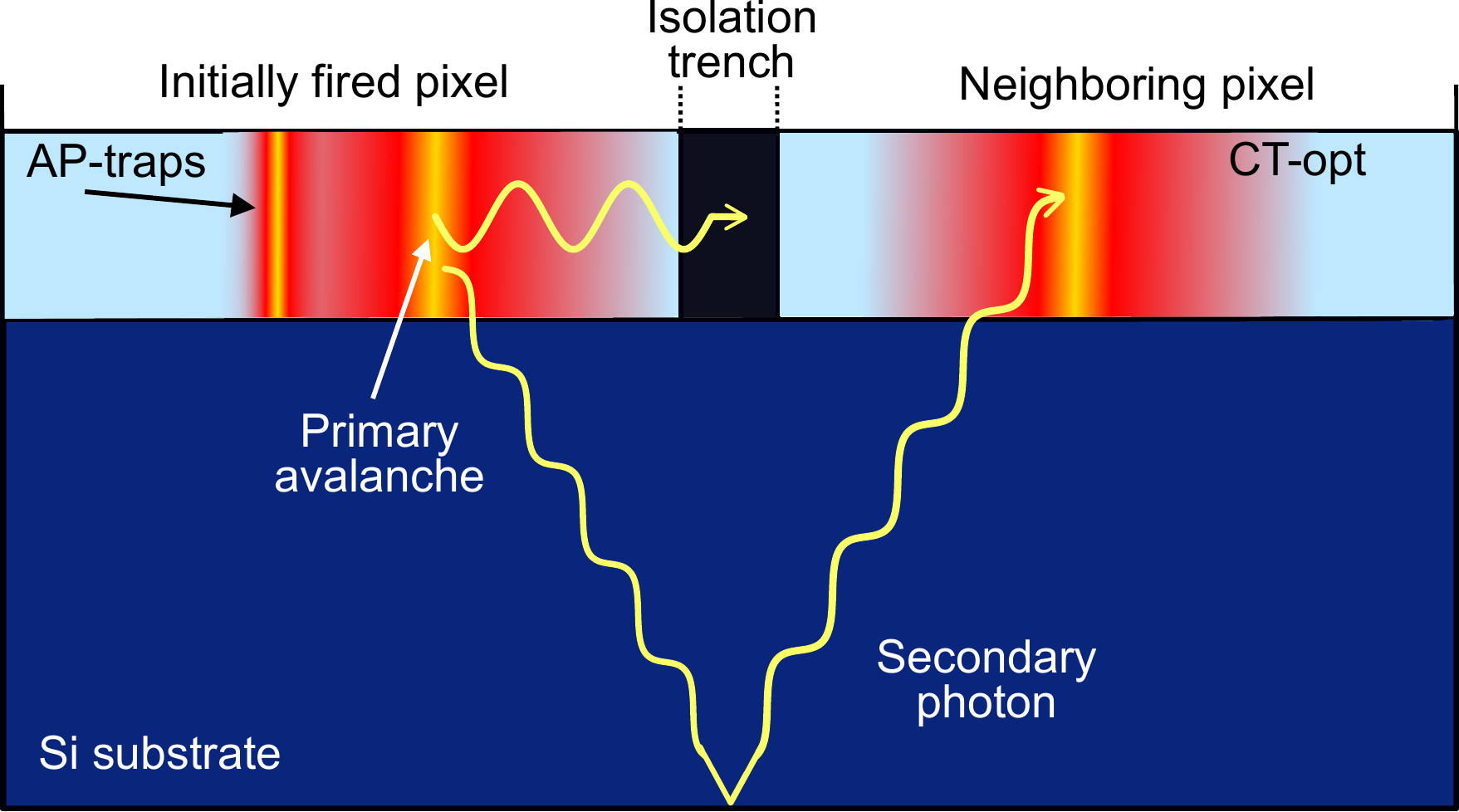}%
\vspace{4 mm}
\includegraphics[width=.7\linewidth]{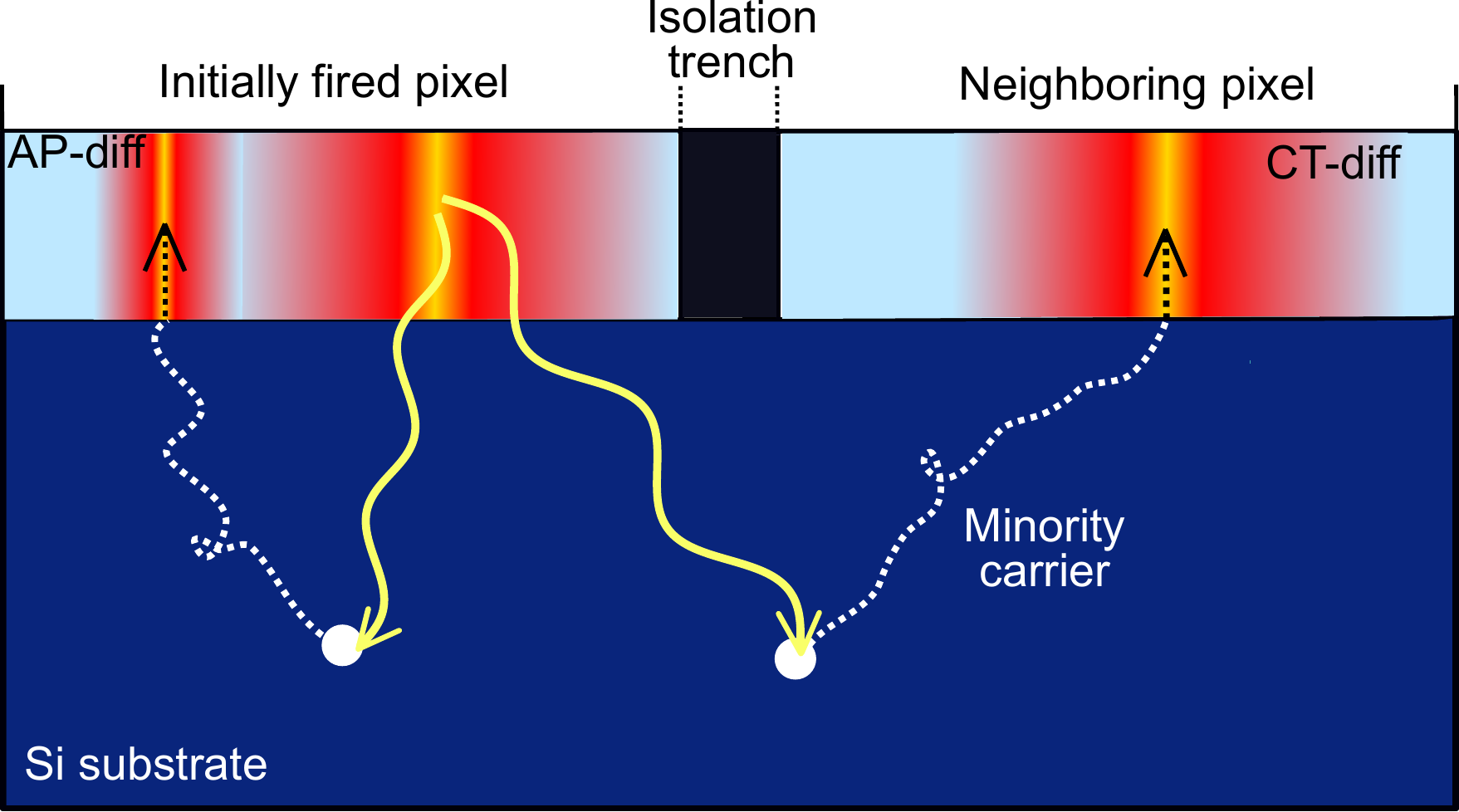}%
\caption{%
Physical processes considered in the analysis. Top: Afterpulsing due to carriers trapped in crystal defects within a
pixel, and almost instantaneous optical crosstalk due to secondary photons reaching neighboring pixels. Bottom:
Secondary photons generate minority carriers in the silicon substrate that may diffuse back to either the primary pixel
or a neighboring one, inducing respectively afterpulsing or delayed crosstalk.%
} \label{fig:processes}
\end{figure}

The physical processes considered in our analysis are illustrated in figure \ref{fig:processes}. In the first place, a
charge carrier of a primary avalanche can be trapped in a crystal defect within the pixel, then when the avalanche has
already been quenched, the carrier may be released triggering a new avalanche. We refer to this process as afterpulsing
due to traps (AP-traps) and it has as many components as types of traps in the silicon lattice. In the second place,
secondary photons of the primary avalanche may reach neighboring pixels triggering almost simultaneously other
avalanches. This is the so-called optical crosstalk (CT-opt). Some SiPMs incorporate trenches filled with an optical
absorber surrounding pixels to prevent this effect, but CT-opt can still be produced by photons reflected on the bottom
surface of the detector. Finally, minority carriers generated in the silicon substrate by secondary photons of the
primary avalanche can diffuse back to either the primary pixel or to a neighboring one, contributing respectively to
crosstalk (CT-diff) or afterpulsing (AP-diff).\footnote{Only the minority carriers are accelerated by the electric
field and so able to trigger an avalanche when they reach a pixel.} The CT-diff is usually referred to as delayed
crosstalk because of the relatively long diffusion times, in contrast to the CT-opt, which is also called prompt
crosstalk.

We presented in \cite{Gallego} an experimental method and an analytical model to describe the CT-opt in a SiPM
fabricated by Hamamatsu. In the present work, our study has been extended with new measurements for the same SiPM as
well as for a second SiPM, also from Hamamatsu. In addition, we have constructed a statistical model accounting for
AP-traps, CT-diff and AP-diff, and preliminary measurements to validate it are shown. A similar analysis of
afterpulsing and CT-diff in a STMicroelectronics SiPM has recently been presented in \cite{Nagy}.

\section{Experimental method}
\label{sec:method}

The experimental method has been described in detail in \cite{Gallego}, and only an overview is given here. The setup
consists of the SiPM with the associated bias circuit, a fast amplifier, and a digital oscilloscope to register and
store the signals for later analysis. Two SiPMs from Hamamatsu were tested: the S10362-11-100C (the one previously used
in \cite{Gallego}) and the S10362-33-100C. Both detectors have a pixel pitch of 100~$\mu$m and basically the same
characteristics, but the first SiPM has 100 pixels while the second one has 900 pixels. Measurements were performed at
dark conditions, that is, an only primary pixel is expected to be triggered at a time. So far, we have characterized
the intrinsic crosstalk and afterpulsing (i.e., no scintillator was attached to the detector) at room temperature.

An algorithm has been developed to perform a detailed waveform analysis of the recorded signals. This dedicated
software calculates numerically the deconvolution of the signal with a negative exponential function that describes the
decay of pulses (time constant of 23~ns for both detectors). This way, pulses are identified as distinct peaks in the
deconvolved signal even if they overlap in the original signal. The position and height of the deconvolution peak
provide us relative measures of the arrival time and amplitude of each pulse. The algorithm resolves and measures the
parameters of pulses as close as 6~ns. In addition, events with two or more pulses with smaller time differences (down
to $\sim 1$~ns) are identified by analyzing the shape and width at threshold level of the deconvolved signal. This
allowed us to characterize separately the nearly instantaneous CT-opt and the other phenomena that result in delayed
secondary pulses. A more precise measurement of the pulse amplitude is obtained as the pulse height with baseline
substraction. To do that, the software performs an exponential extrapolation of the original signal in a small region
prior to each pulse.

\section{Optical crosstalk}
\label{sec:crosstalk}

The CT-opt gives rise to pulses with amplitudes two or more times higher than that corresponding to only one triggered
pixel. The relative intensities of the pulse amplitude spectrum measured at dark conditions thus provides the
probability distribution $P(k)$ of the total number $k$ of simultaneously triggered pixels per primary avalanche, i.e.,
$k-1$ CT-opt excitations. We produced this spectrum using the pulse height with baseline substraction to optimize the
resolution in $k$. In addition, quality cuts were applied to reject low-amplitude pulses due to pixels that were
triggered while being recharged after the last avalanche breakdown.

We presented in \cite{Gallego} a statistical model based on the assumption that secondary photons can only reach a
certain neighborhood of pixels around the primary one. Cascades of CT-opt excitations propagating through the whole
array of pixels were included, but taking into account that pixels that have already been triggered cannot be triggered
twice. The $P(k)$ distribution was then calculated by identifying all the CT-opt ``histories'' that contribute to each
probability considering four possible arrangements of neighbors: the 4 nearest neighbors, the 8 nearest neighbors, the
8 L-connected neighbors and all neighbors (see \cite{Gallego} for details). For simplicity, we assumed the same CT-opt
probability for any neighbor and ignored the boundaries of the array of pixels for the calculations. Analytical
expressions for the mean and variance of the $P(k)$ distribution were also obtained.

\begin{figure}[t!]
\centering%
\includegraphics[width=.7\linewidth]{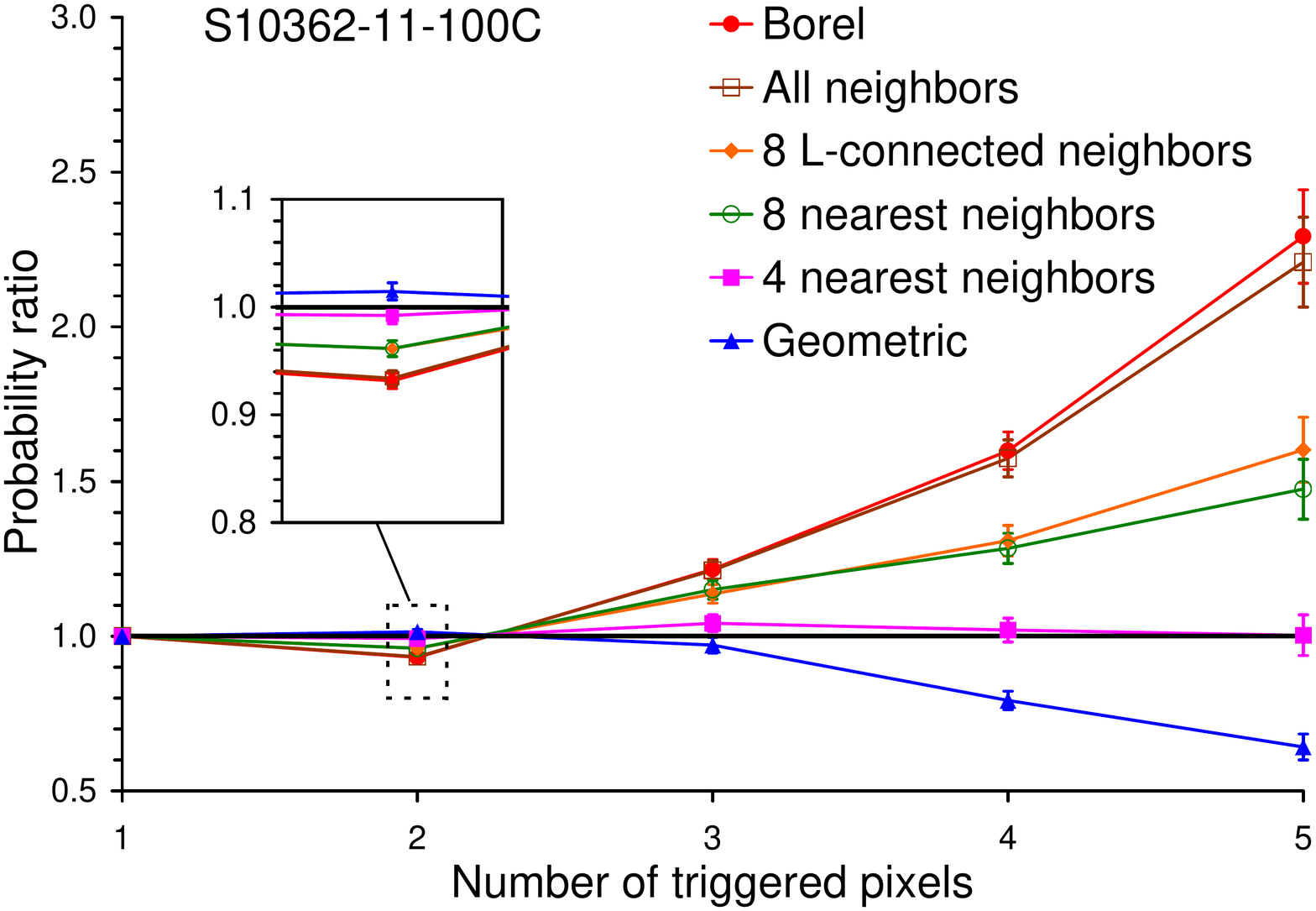}%

\includegraphics[width=.7\linewidth]{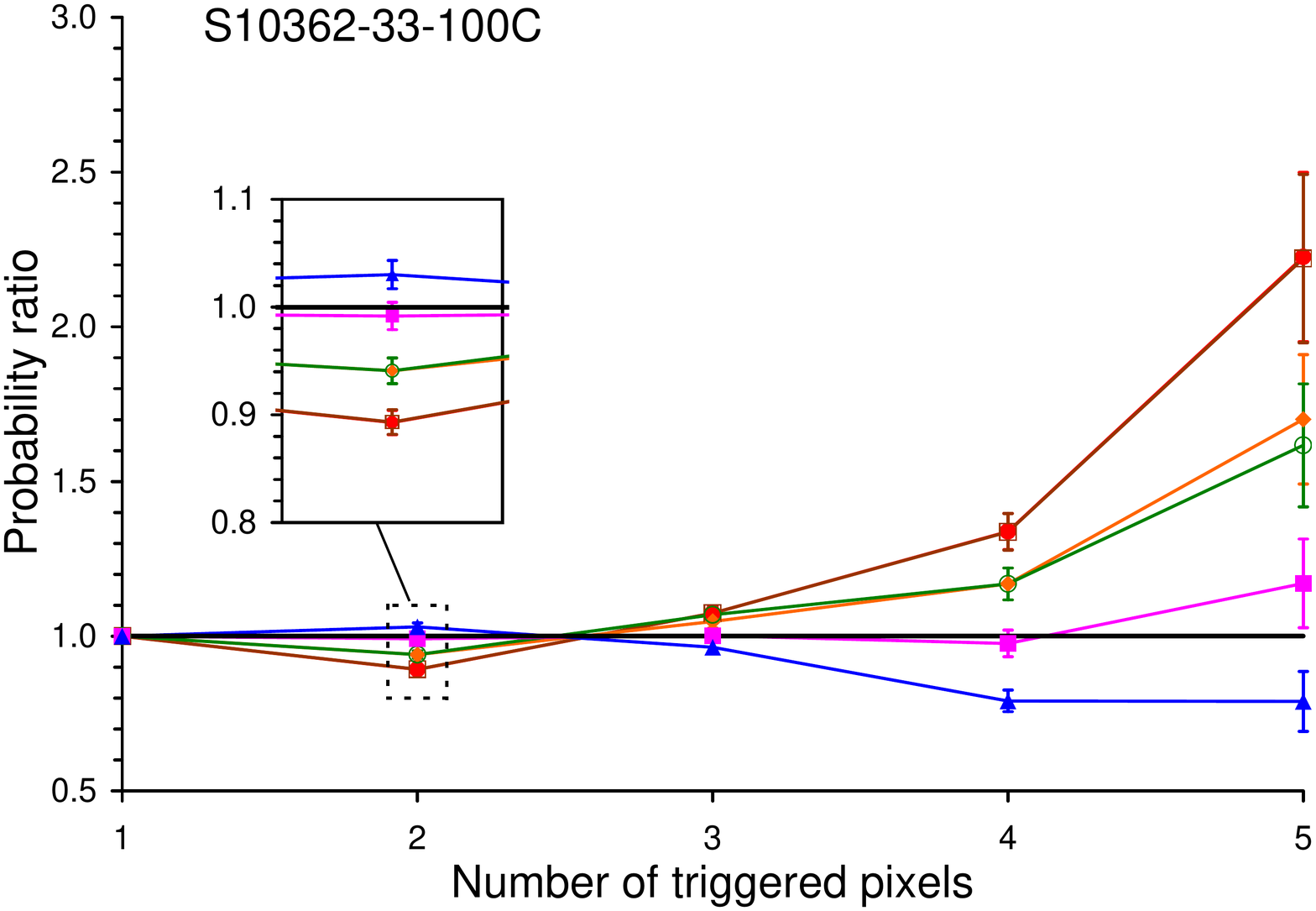}%
\caption{%
Ratio of model predictions of $P(k)$ for optical crosstalk over the experimental ones measured at dark conditions and a
given bias voltage. Results for the two tested detectors are consistent with the hypothesis of the 4 nearest
neighbors. See text for details.%
} \label{fig:crosstalk}
\end{figure}

In figure \ref{fig:crosstalk}, the model predictions for the four arrangements of neighbors are compared with the
experimental probabilities obtained at a given bias voltage for both the S10362-11-100C detector (top plot, taken from
\cite{Gallego}) and the S10362-33-100C detector (bottom plot). In particular, the probability ratio (theoretical over
experimental) is represented versus $k$ from 1 to 5, where the model predictions were normalized so that the first
probability $P(1)$ matches the experimental one, i.e., their ratio is unity. The error bars of the remaining ratios
were determined propagating the uncertainties of the experimental probabilities. The comparison includes two previous
analytical models described in \cite{Vinogradov2}, where $P(k)$ is assumed to follow a geometric or a Borel
distribution. As explained in \cite{Gallego}, these two models can be regarded as limit situations of ours.

\begin{figure}[t!]
\centering%
\includegraphics[width=.7\linewidth]{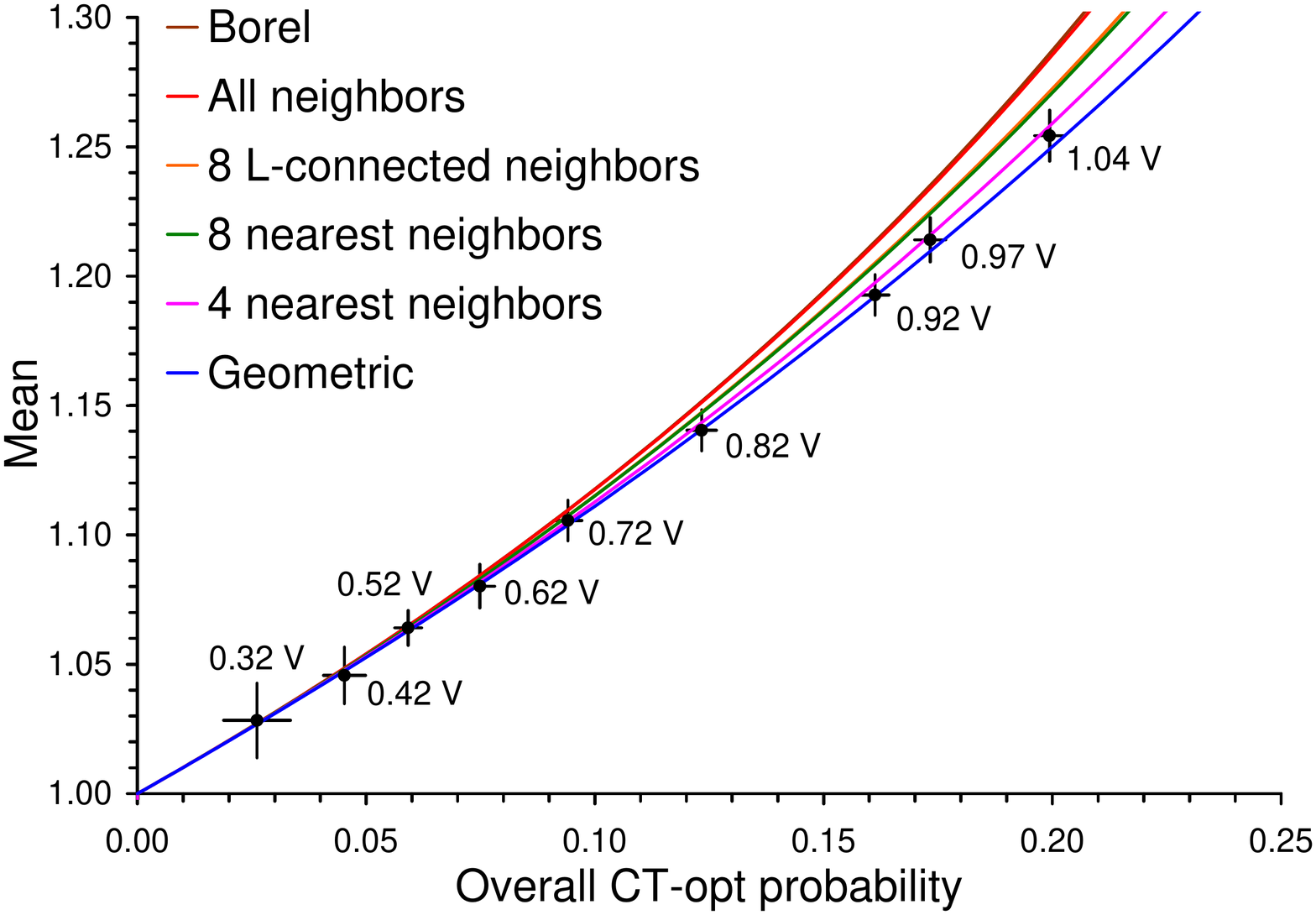}%

\includegraphics[width=.7\linewidth]{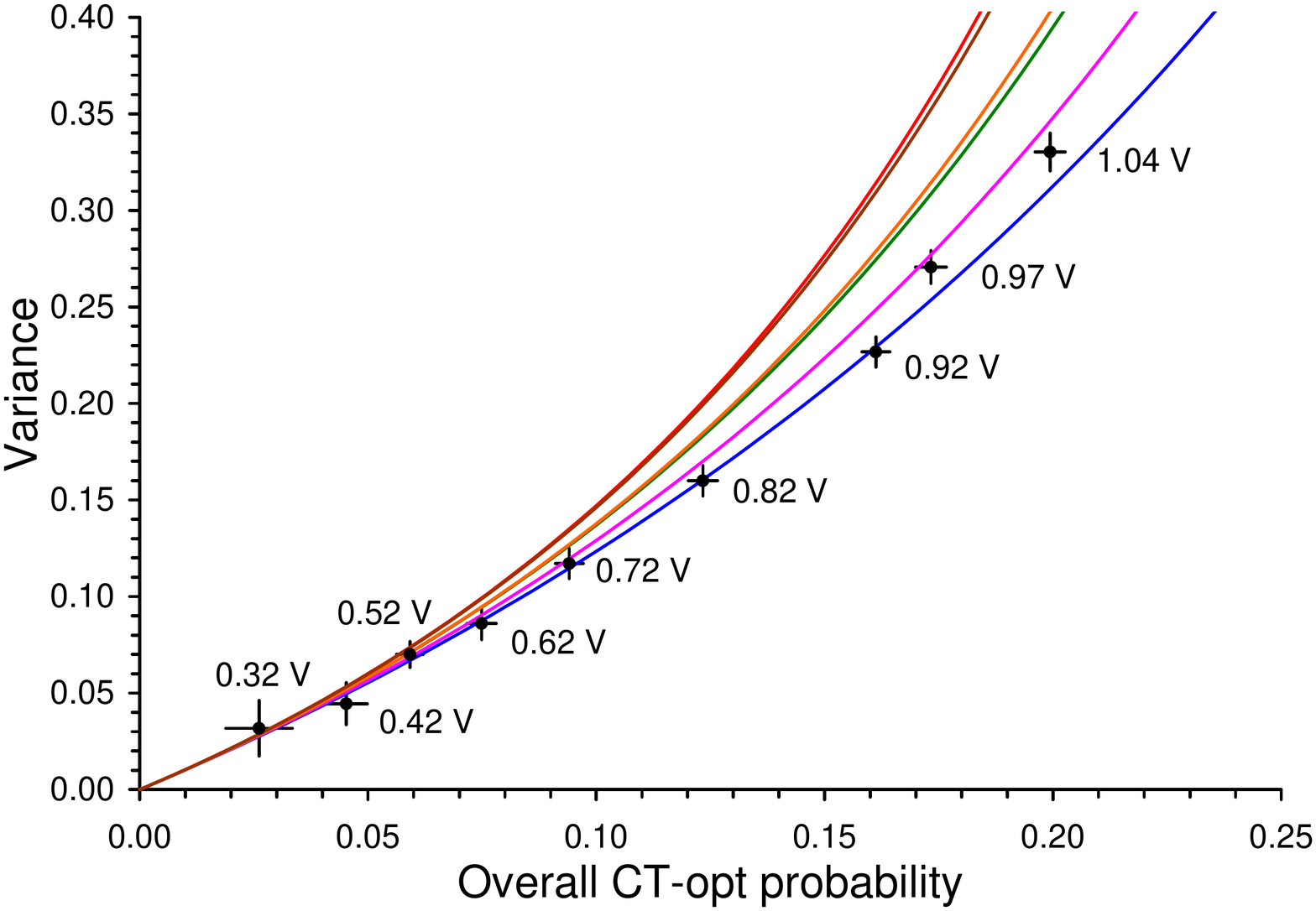}%
\caption{%
Comparison of the model predictions (lines) with experimental data (points) of optical crosstalk taken at different
bias voltages. Results are for the mean (top) and the variance (bottom) of the $P(k)$ distribution for the
S10362-11-100C detector. The legend shows the order from top to down of the lines representing the model predictions.%
} \label{fig:crosstalk_V}
\end{figure}

In figure \ref{fig:crosstalk_V}, the model predictions of the mean and variance of the $P(k)$ distribution are compared
with new experimental results for the S10362-11-100C detector as a function of the overall CT-opt probability, defined
as $1-P(1)$. The overvoltage, i.e., bias voltage minus breakdown voltage, at which measurements were performed is also
indicated for each data point (the overall CT-opt probability grows quadratically with overvoltage).

The comparisons shown in figures \ref{fig:crosstalk} and \ref{fig:crosstalk_V} reveal that the only model consistent
with our experimental data is the one assuming that CT-opt solely takes place in the 4 nearest neighbors of each pixel.
This is the case for both tested detectors, which have different array sizes. Therefore, it can also be concluded that
border effects are unimportant to describe CT-opt effects at these conditions.

\section{Afterpulsing and delayed crosstalk}
\label{sec:afterpulsing}

At dark conditions, the distribution of the pulse arrival time measured with respect to the previous pulse is a
combination of the contributions of both uncorrelated dark noise and correlated secondary pulses (i.e., AP-traps,
AP-diff and CT-diff). Since we were interested in measuring the probability of secondary pulses per primary avalanche,
we obtained this time distribution selecting pairs of pulses in which the first one (the primary) has an amplitude
corresponding to one only triggered pixel, that is, without CT-opt. In addition, this first pulse of the pair was
required to be at least 500~ns far from previous pulses to avoid mixing contributions of secondary pulses from
different primaries.

Assuming that the total number of secondary pulses of each type per primary avalanche is Poisson distributed, the above
time distribution can be expressed as

\begin{align}
p(t)\cdot{\rm d}t=&\exp\left[-R_{\rm DC}\cdot(t-t_{\rm min})-\sum_i\lambda_i\cdot\int_{t_{\rm min}}^t f_i(s)\cdot{\rm d}s\right] \nonumber\\
&\left[R_{\rm DC}+\sum_i\lambda_i\cdot f_i(t)\right]\cdot{\rm d}t\,,
\label{eq:afterpulsing}
\end{align}
where $t_{\rm min}=10$~ns is the minimum time difference allowed between pulses due to analysis limitations, $R_{\rm
DC}$ is the dark count rate, $\lambda_i$ is the mean number of secondary pulses of type $i$ per primary avalanche, and
$f_i(t)$ is the corresponding normalized time distribution. The details of this model, in particular the
parameterization of the $f_i(t)$ functions, are still under progress and will be explained in a paper in preparation.
Only the general aspects are given below.

\begin{figure}[t]
\centering
\includegraphics[width=.7\linewidth]{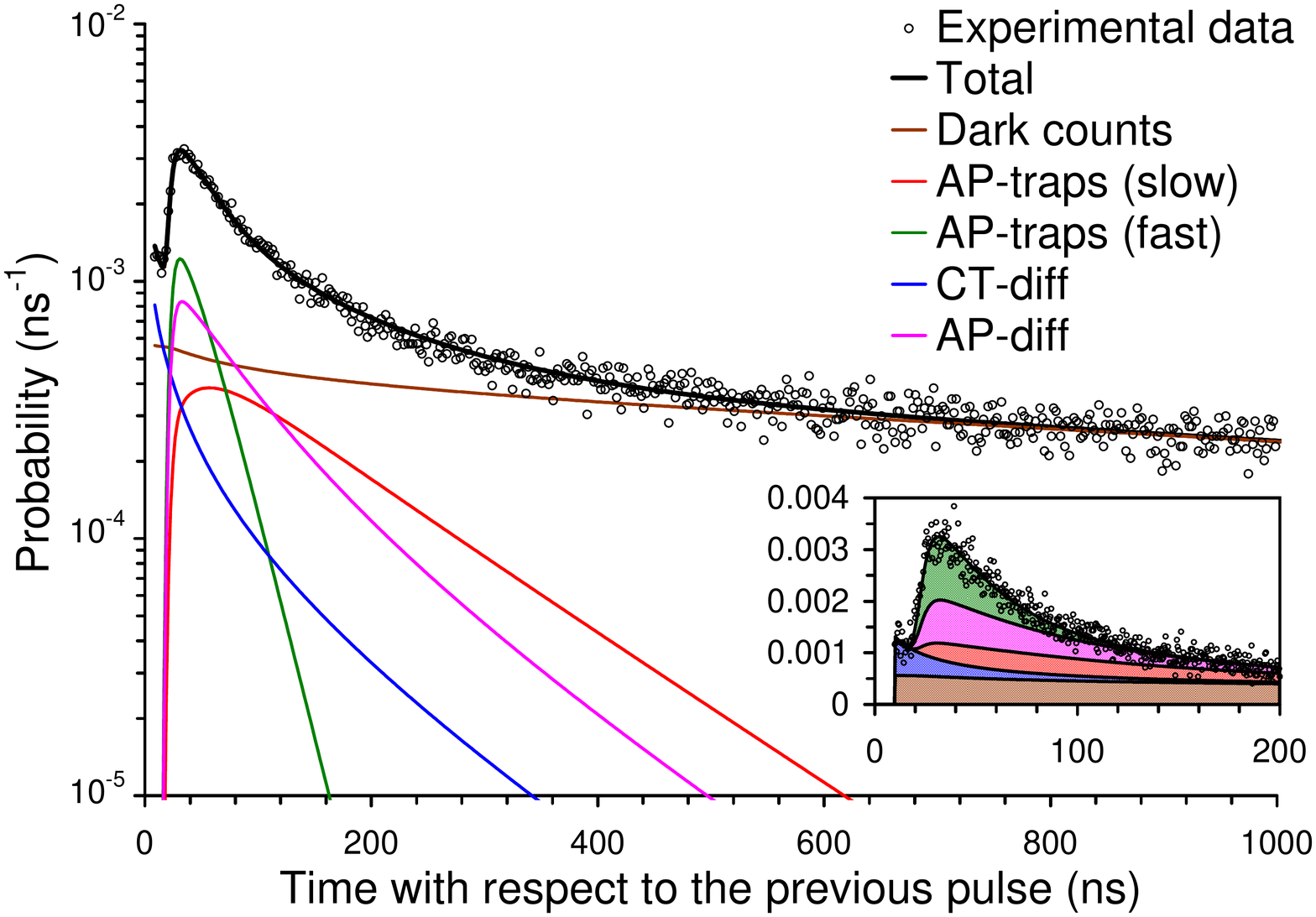}
\caption{%
Fit of the model of afterpulsing and delayed crosstalk to the experimental time distribution of secondary pulses for
the S10362-11-100C detector. The model predictions for each component of secondary pulses are shown. The accumulated
contributions of the components at $t<200$~ns are also shown in the inset (linear scale).%
} \label{fig:afterpulsing}
\end{figure}

The $f(t)$ distribution for AP-traps was characterized by the product of three factors: a negative exponential
distribution describing the release of trapped carriers; the exponential pixel recharging after the primary avalanche,
which is proportional to both the gain and the probability of the secondary avalanche; and the $t$-dependent fraction
of afterpulses (of any type) with amplitudes above the detection threshold applied in the analysis software. The last
two factors, which are also applicable to the $f(t)$ distribution for AP-diff, account for the steep break of the
experimental time distribution between 20 and 30~ns (see figure \ref{fig:afterpulsing}) and were determined from the
time-amplitude information of afterpulses with short delay.\footnote{Afterpulses with short delay have smaller
amplitudes as a consequence of the pixel recharging, and thus, they can be distinguished from dark counts and CT-diff
pulses produced by fully recharged pixels.}

A Monte Carlo simulation was carried out to describe the timing and relative contributions of the AP-diff and CT-diff
effects (bottom plot of figure \ref{fig:processes}). Values for the physical parameters related to the emission and
absorption of secondary photons, the diffusion of minority carriers and the layer configuration of the chip were varied
within a wide interval around typical data taken from the literature (e.g., see \cite{Otte}). Preliminary simulation
results showed that the transient time distribution can be properly described by the product of the exponential decay
of the number of minority carriers due to recombination mechanisms and a negative power law. Also, the relative
contributions of AP-diff and CT-diff were found to be strongly dependent on the pixel pitch. For instance, for a very
small pixel pitch of 10~$\mu$m, the AP-diff component is negligible compared to the CT-diff one, while both components
are significant for a pixel pitch of 100~$\mu$m (present case).

A fit of our model to experimental data for the S10362-11-100C detector at an overvoltage of $\sim 1$~V is shown in
figure \ref{fig:afterpulsing}, where the contributions of the different types of secondary pulses and of dark counts
are represented by thin lines. The corresponding accumulated contributions are also shown in the inset of the figure in
a linear scale. This result corresponds to a given set of parameters assumed in the above simulation, which involves
that the $f(t)$ distributions and the relative contributions of AP-diff and CT-diff are fixed (i.e., only one $\lambda$
parameter for both components is let free in the fit). In this case, two AP-traps components with different mean
release time $\tau$ were necessary to describe the data: a fast component with $\tau_{\rm fast}\approx 25$~ns and a
slow one with $\tau_{\rm slow}\approx150$~ns. On the other hand, when letting all the parameters related to AP-diff and
CT-diff free, good quality fits were also obtained even omitting any AP-traps component. Similar results were obtained
for the S10362-33-100C detector (not shown here). We are making the simulation more realistic in order to be more
restrictive on the fitting parameters.

\begin{figure}[t]
\centering
\includegraphics[width=0.7\linewidth]{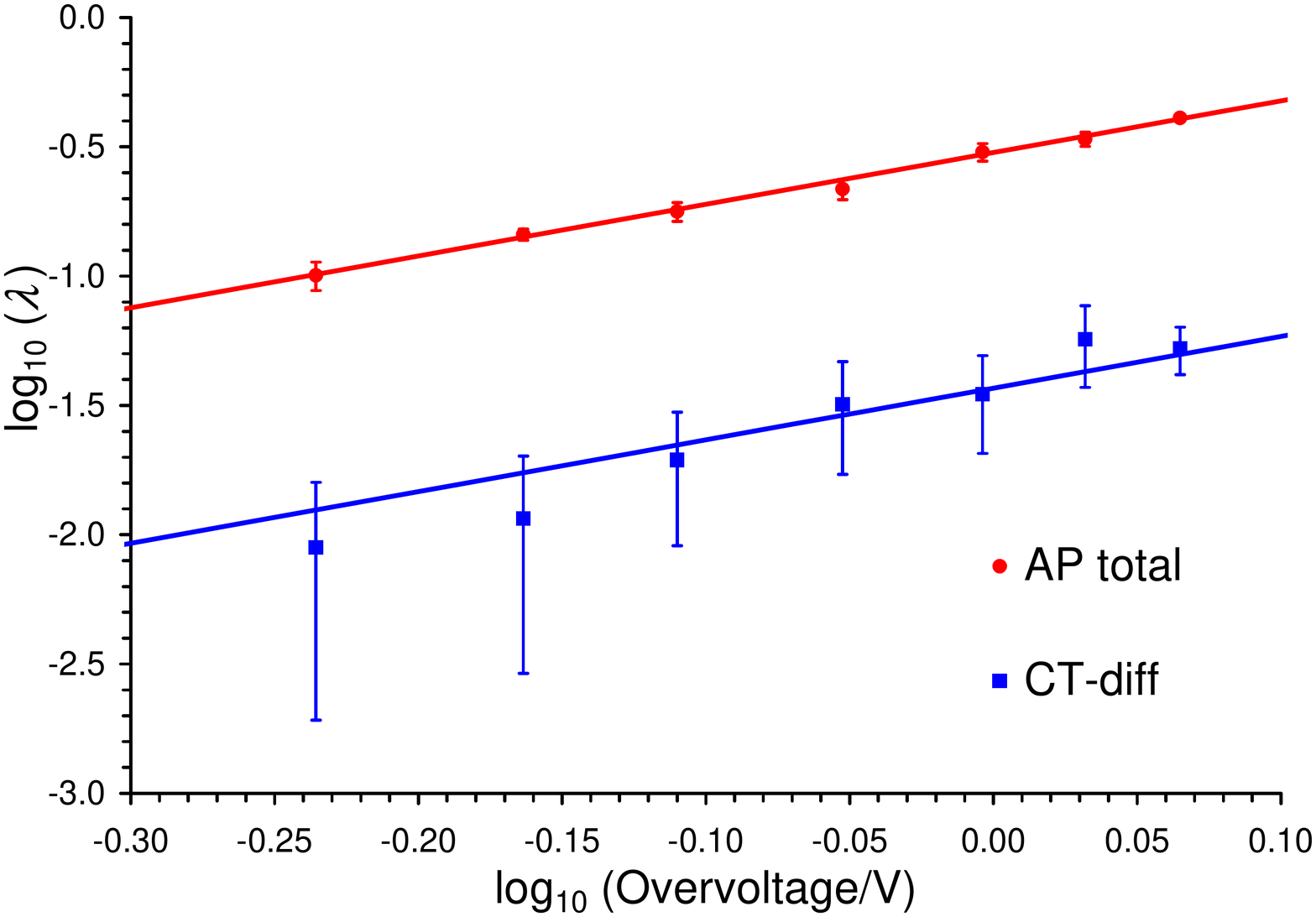}
\caption{%
Average number of afterpulses (of any type) and pulses due to delayed crosstalk per primary avalanche as a function of
overvoltage for the S10362-11-100C detector. The lines represent the best fits of quadratic functions of type $y=a\cdot
x^2$. The uncertainties are statistical only.%
} \label{fig:afterpulsing_V}
\end{figure}

Even though the number of AP components has not been determined yet, the sum of the $\lambda$ parameters of all the AP
components was found to be nearly independent of the assumptions made on the fit. The $\lambda$ parameter associated to
the CT-diff was roughly fit independent as well, because it is basically determined by data at $t\lesssim20$~ns, where
AP contributions can be neglected. These $\lambda$ parameters, i.e., the one for the total contribution of afterpulsing
and the one for CT-diff, are represented in figure \ref{fig:afterpulsing_V} as a function of overvoltage, showing a
pure quadratic behavior, as expected.\footnote{These results were obtained by fitting the $\lambda$ parameters, but
using the $f(t)$ distributions determined in the fit shown in figure \protect\ref{fig:afterpulsing} at a given bias
voltage. Appropriate corrections were applied to account for the overvoltage dependence of the detection threshold
effects on the afterpulsing measurements.}

\section{Conclusions}
\label{sec:conclusions}

We have developed an experimental method based on a waveform analysis to characterize the crosstalk and afterpulsing in
SiPMs. Results were obtained for the S10362-11-100C and S10362-33-100C detectors from Hamamatsu at dark conditions and
room temperature.

Measurements of optical crosstalk have been compared with an analytical model that we presented in a previous
publication. Our data for both detectors are consistent with the hypothesis that optical crosstalk is only possible
between adjacent pixels.

A statistical model of afterpulsing and delayed crosstalk has been constructed including a detailed parameterization of
the time distributions of the different types of secondary pulses. Our preliminary results show a significant
probability of afterpulsing in both detectors, although the characterization of its components is still under study.
The delayed crosstalk was also found to be important, mostly at short time.

These studies are being extended to characterize other SiPMs and for a larger variety of conditions. We intend to apply
our models to describe crosstalk and afterpulsing effects on particular experimental cases.

\section*{Acknowledgments}
\label{sec:acknoledgments}

This work was supported by MINECO (FPA2012-39489-C04-02) and CONSOLIDER CPAN CSD2007-42.

\end{document}